\documentclass[aps,prd,nofootinbib,onecolumn,preprintnumbers,floatfix,superscriptaddress,bibliography]{revtex4-2}
\usepackage[colorlinks, linkcolor=red, anchorcolor=blue, citecolor=blue, colorlinks=true, urlcolor=blue]{hyperref}
\usepackage{amsmath}
\usepackage{graphicx}
\usepackage{fontawesome}
\usepackage{ulem}
\usepackage{bm}
\usepackage{xspace}
\usepackage{amssymb}
\usepackage{gensymb}
\usepackage{multirow}
\usepackage{threeparttable}
\usepackage{amsthm}
\usepackage{mathrsfs}
\usepackage{indentfirst}
\usepackage{subfigure}
\usepackage[figuresright]{rotating}
\usepackage{epstopdf}
\usepackage{breakurl}
\usepackage{lipsum}
\usepackage{cancel}
\usepackage{slashed}

\begin{document}

\title{Axion effects on gamma-ray spectral irregularities. II: Implications of EBL absorption}

\author{Hai-Jun Li} 
\email{lihaijun@itp.ac.cn}
\affiliation{Key Laboratory of Theoretical Physics, Institute of Theoretical Physics, Chinese Academy of Sciences, Beijing 100190, China}
 
\author{Wei Chao} 
\email{chaowei@bnu.edu.cn}
\affiliation{Center for Advanced Quantum Studies, School of Physics and Astronomy, Beijing Normal University, Beijing 100875, China} 
\affiliation{Key Laboratory of Multi-scale Spin Physics, Ministry of Education, Beijing Normal University, Beijing 100875, China}
 
\author{Xiu-Hui Tan} 
\email{tanxh@itp.ac.cn}
\affiliation{Key Laboratory of Theoretical Physics, Institute of Theoretical Physics, Chinese Academy of Sciences, Beijing 100190, China}
 
\author{Yu-Feng Zhou}
\email{yfzhou@itp.ac.cn}
\affiliation{Key Laboratory of Theoretical Physics, Institute of Theoretical Physics, Chinese Academy of Sciences, Beijing 100190, China}
\affiliation{School of Physical Sciences, University of Chinese Academy of Sciences, Beijing 100049, China}
\affiliation{School of Fundamental Physics and Mathematical Sciences, Hangzhou Institute for Advanced Study, UCAS, Hangzhou 310024, China}
\affiliation{International Centre for Theoretical Physics Asia-Pacific, Beijing/Hangzhou, China}

\preprint{ITP-24-240, BNU-24-095}

\date{\today}

\begin{abstract}

The extragalactic background light (EBL) plays a crucial role in the propagation of high-energy particles throughout the Universe.
In this work, we explore the impact of the EBL absorption effect on photon to axionlike particle (ALP) conversions from the very-high-energy gamma-ray spectral irregularities.
For our purpose, we select four BL Lac blazars: Markarian\,501, 1ES\,0229+200, PKS\,0301-243, and PKS\,0447-439 for analysis.
Their redshifts range from approximately 0.03 to 0.34.
We first discuss the EBL absorption effect on the gamma-ray spectral energy distributions (SEDs) using three common EBL spectral models: Finke-10, Franceschini-17, and Saldana-Lopez-21.
Then we consider the photon-ALP conversions in astrophysical magnetic fields.
The best-fit chi-square distributions of these EBL models under the ALP assumption in the ALP parameter $\{m_a, g_{a\gamma}\}$ plane are provided, showing similar distributions.
For comparison, we define a new delta chi-square, $\chi_d^2$, to quantify the difference in chi-square values.
The distributions of $\chi_d^2$ and the gamma-ray SEDs corresponding to the maximum delta chi-square, $\chi^2_{d, \rm max}$, are also presented for comparison. 
Our results indicate that the influence of these different EBL models is non-dominant at the low-redshift gamma-ray axionscope.
In these cases, choosing the latest model, Saldana-Lopez-21, is sufficient.
However, as the redshift of the sources increases, this influence becomes more significant.


\end{abstract}
\maketitle

 
\section{Introduction}

Axions are excellent candidates for new physics.
The QCD axion was originally introduced by the Peccei-Quinn (PQ) mechanism to  dynamically solve the strong CP problem in the Standard Model (SM) \cite{Peccei:1977hh, Peccei:1977ur, Weinberg:1977ma, Wilczek:1977pj}, meanwhile, it also provides a natural source for cold dark matter (DM) through the misalignment mechanism \cite{Preskill:1982cy, Abbott:1982af, Dine:1982ah}.
On the other hand, the axionlike particle (ALP), predicted by a variety of theories \cite{Arvanitaki:2009fg, Svrcek:2006yi}, is also the attractive DM candidate \cite{Cadamuro:2011fd, Arias:2012az, Chao:2022blc}, but does not have to solve the strong CP problem.
The axion can couple to the photon with the effective Lagrangian
\begin{eqnarray}
\begin{aligned}
\mathcal{L}&=\dfrac{1}{2}\partial_\mu a\partial^\mu a - \dfrac{1}{2}m_a^2a^2 -\dfrac{1}{4}g_{a\gamma}aF_{\mu\nu}\tilde{F}^{\mu\nu}\\
&=\dfrac{1}{2}\partial_\mu a\partial^\mu a - \dfrac{1}{2}m_a^2a^2 + g_{a\gamma}a\textbf{E}\cdot\textbf{B}\, ,
\end{aligned}
\end{eqnarray}
where $a$ and $m_a$ donate the axion field and axion mass, respectively, $g_{a\gamma}$ is the axion-photon coupling constant, $F_{\mu\nu}$ ($\tilde{F}^{\mu\nu}$) is the (dual) electromagnetic field tensor, $\textbf{E}$ and $\textbf{B}$ are the local electric and magnetic field vectors, respectively.
In the QCD axion scenario, the axion mass $m_a$ and coupling $g_{a\gamma}$ are interrelated, whereas in the ALP scenario, they are considered independent parameters.
Therefore, the ALP has a much wider $\{m_a, g_{a\gamma}\}$ parameter space than the QCD axion.
See ref.~\cite{ciaran_o_hare_2020_3932430} for the latest ALP-photon limits.

The coupling between the ALPs and very-high-energy (VHE; $\sim \mathcal{O}(100)\, \rm GeV$) photons in astrophysical magnetic fields could lead to some detectable signals, such as a reduced TeV opacity of the Universe \cite{DeAngelis:2007dqd, Mirizzi:2007hr, Simet:2007sa, Hooper:2007bq}.
The VHE gamma-rays from the extragalactic sources, such as blazars, are mainly affected by the extragalactic background light (EBL) absorption effect due to the electron-positron pair production process
\begin{eqnarray}
\gamma_{\rm TeV} + \gamma_{\rm EBL} \to e^- + e^+\, ,
\label{pair_production}
\end{eqnarray}
where $\gamma_{\rm TeV}$ and $\gamma_{\rm EBL}$ are the VHE photons and background photons, respectively.
By taking into account the photon-ALP conversions and back-conversions in the simulated astrophysical magnetic fields, the EBL absorption effect can be mitigated, resulting in the Universe that is potentially more transparent than previously thought solely based on the EBL absorption \cite{HESS:2007xak, MAGIC:2008sib}. 
Furthermore, it also offers a natural mechanism for constraining ALP properties, and many similar studies have recently been conducted within this axionscope scenario \cite{Abramowski:2013oea, TheFermi-LAT:2016zue, Kohri:2017ljt, Zhang:2018wpc, Ivanov:2018byi, Liang:2018mqm, Libanov:2019fzq, Long:2019nrz, Guo:2020kiq, Bi:2020ths, Li:2020pcn, Cheng:2020bhr, Li:2021gxs, Li:2021zms, Dessert:2022yqq, Li:2022jgi, Jacobsen:2022swa, Mastrototaro:2022kpt, Li:2022mcf, Noordhuis:2022ljw, Pant:2022ibi, Gao:2023dvn, Pant:2023omy, Pant:2023lnz, Li:2023qyr, Li:2024ivs, Guo:2024oqo, Li:2024zst, Ruz:2024gkl, Porras-Bedmar:2024uql, Gao:2024wpn, BetancourtKamenetskaia:2024gcv, Zhu:2024kmu}.

Previous studies show that the source magnetic field parameters, such as the strength of the core magnetic field, have the most significant influence on the limits of ALP properties \cite{Li:2020pcn, Li:2021gxs}.
Moreover, the source redshift uncertainty has an impact, both the underestimated and overestimated redshifts can affect the ALP limits \cite{Li:2022pqa}.
Additionally, the EBL absorption effect itself, more precisely, the EBL spectral models, may also have an impact on the final results.
However, it can be naively predicted that this effect could be much smaller than those mentioned above, but it is worth investigating them quantitatively.

\begin{table}[b]
\centering
\caption{The redshift and position information of the four BL Lac blazars: Markarian\,501, 1ES\,0229+200, PKS\,0301-243, and PKS\,0447-439.
See \url{http://tevcat2.uchicago.edu} for more details.
The sources of the gamma-ray data used in this work are also listed.}
\begin{ruledtabular}
\begin{tabular}{lcccc}
Source                  &  redshift    & Gal.Long.\,(deg) & Gal.Lat.\,(deg) &  gamma-ray data\\
\hline
Markarian\,501      &  0.034      &   63.60     &  38.86    &  Fermi-LAT+HAWC \cite{HAWC:2021obx} \\
1ES\,0229+200     &  0.1396    &  152.97    &  -36.61   &   Fermi-LAT+MAGIC \cite{MAGIC:2022piy} \\
PKS\,0301-243      &  0.2657    &   214.63   &   -60.19  & Fermi-LAT+H.E.S.S. \cite{HESS:2013uta} \\
PKS\,0447-439      &  0.343      &   248.81   &   -39.91  &  Fermi-LAT+H.E.S.S. \cite{HESS:2013mfa} \\
\end{tabular}
\end{ruledtabular}
\label{tab_1}
\end{table}

In this work, we explore the impact of the EBL absorption effect on photon-ALP conversions from the VHE gamma-ray spectral irregularities.
In this regard, the gamma-ray source should be selected with a relatively certain redshift.
Meanwhile, we should choose different sources for analysis, and the redshift of these sources must vary from low to high.
For our purpose, we select four BL Lac blazars: Markarian\,501 (with the redshift $z_0=0.034$), 1ES\,0229+200 (with the redshift $z_0=0.1396$), PKS\,0301-243 (with the redshift $z_0=0.2657$), and PKS\,0447-439 (with the redshift $z_0=0.343$).
Their redshifts vary from $\sim0.03-0.34$, it may be a good choice for us to explore the impact of the EBL absorption effect with these sources.
Here we use the relatively latest gamma-ray data of these sources for analysis.
See also table~\ref{tab_1} for their position information and the sources of the gamma-ray data.
On the other hand, the EBL spectral model should be chosen as a relatively up-to-date one, and there should be certain distinctions among the energy spectra of these models.
Therefore, we select three EBL models: Finke-10 \cite{Finke:2009xi}, Franceschini-17 \cite{Franceschini:2017iwq}, and Saldana-Lopez-21 \cite{Saldana-Lopez:2020qzx}.
The first EBL spectral model exhibits certain differences from the second and third in the far-infrared regions. 
Both the second and third models represent the latest EBL models, and the observations for the third model originate from space, offering greater reliability compared to those conducted from the Earth.
We first discuss the EBL absorption effect on the gamma-ray spectral energy distributions (SEDs) with these three EBL spectral models.
Then we consider the photon-ALP conversions in astrophysical magnetic fields.
The best-fit chi-square distributions of these EBL models under the ALP assumption in the ALP parameter $\{m_a, g_{a\gamma}\}$ plane are given, showing similar distributions.
For comparison, we define a new delta chi-square $\chi_d^2$ to quantify the chi-square difference.
The distributions of $\chi_d^2$ and the gamma-ray SEDs corresponding to the maximum delta chi-square $\chi^2_{d, \rm max}$ in the $\{m_a, g_{a\gamma}\}$ plane are also shown.
Finally, we find that there is only a minor influence from the different EBL models at the low-redshift gamma-ray axionscope. 
However, as the redshift of the sources increases, this impact becomes more pronounced, which can be observed from the gamma-ray SEDs. 
On the other hand, the uncertainty in the observed spectra can also directly affect the results. 
Overall, for the investigation of photon-ALP coupling from the low-redshift gamma-ray sources, choosing the EBL spectral model Saldana-Lopez-21 is sufficient.

The rest of this paper is structured as follows.
In section~\ref{sec_null}, we introduce the EBL absorption effect on the VHE gamma-ray propagation and show the gamma-ray data of the selected blazars.
In section~\ref{sec_ALP}, we discuss the photon-ALP conversions in astrophysical magnetic fields.
The analysis and results under the ALP assumption are given in section~\ref{sec_results}.
Finally, the conclusion is given in section~\ref{sec_onclusion}.

\section{The gamma-ray SEDs under the null hypothesis}
\label{sec_null}

In this section, we first introduce the EBL absorption effect on the VHE gamma-ray propagation, then we show the gamma-ray SEDs of the selected BL Lac blazars Markarian\,501, 1ES\,0229+200, PKS\,0301-243, and PKS\,0447-439.

\subsection{EBL absorption effect}
\label{EBL_absorption}

The EBL plays a crucial role in determining the opacity of cosmic high-energy photons and acts as a constraint on the propagation of high-energy particles throughout the Universe.
In general, due to eq.~(\ref{pair_production}) the main effect on the VHE photons (with the high-energy $E$) from the extragalactic space is the EBL photons (with the low-energy $\omega$) absorption effect with the absorption factor $e^{-\tau}$
\begin{eqnarray}
\Phi(E)=e^{-\tau}\Phi_{\rm int}(E)\, ,
\end{eqnarray}
where $\Phi(E)$ is the gamma-ray expected spectrum, $\Phi_{\rm int}(E)$ is the intrinsic spectrum, and $\tau$ is the optical depth. 
This optical depth can be described by \cite{Franceschini:2008tp}
\begin{eqnarray}
\tau=c \int_0^{z_0} \dfrac{{\rm d}z}{(1+z)H(z)}\int_{E_{\rm th}}^{\infty}{\rm d}\omega\dfrac{{\rm d}n(z)}{{\rm d}\omega} \bar{\sigma}(E,\omega,z)\, ,
\end{eqnarray}
with the Hubble expansion rate
\begin{eqnarray}
H(z)=H_0\sqrt{\left(1+z\right)^2\left(1+\Omega_m z\right)-z\left(2+z\right)\Omega_\Lambda}\, ,
\end{eqnarray}
where $z_0$ is the source redshift, $E_{\rm th}$ is the threshold energy, $\bar{\sigma}(E,\omega,z)$ is the integral pair-production cross section, ${\rm d}n(z)/{\rm d}\omega$ is the EBL proper number density, $H_0 \simeq 67.4\, \rm km\, s^{-1} \,Mpc^{-1}$, $\Omega_m \simeq 0.315$, and $\Omega_\Lambda \simeq 0.685$ \cite{ParticleDataGroup:2024cfk}.

\begin{figure}[t]
\centering
\includegraphics[width=0.65\textwidth]{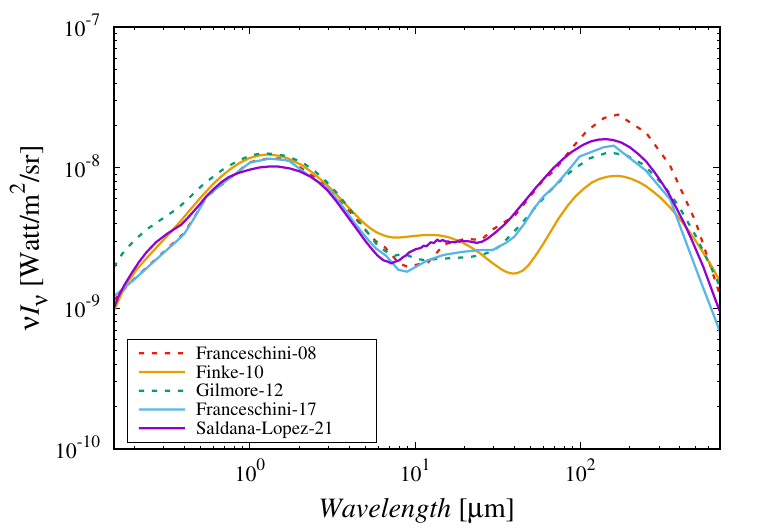}  
\caption{The EBL spectral models.
The lines in different colors represent the spectral models: Franceschini-08 \cite{Franceschini:2008tp}, Finke-10 \cite{Finke:2009xi}, Gilmore-12 \cite{Gilmore:2011ks}, Franceschini-17 \cite{Franceschini:2017iwq}, and Saldana-Lopez-21 \cite{Saldana-Lopez:2020qzx}, respectively.
The solid lines represent the three spectral models used in this work.}
\label{fig_ebl_model}
\end{figure} 

In our previous studies, we did not consider different EBL spectral models, but instead opted for a common one. 
However, in this work, we will take into account the impact of various EBL models.
See figure~\ref{fig_ebl_model} for several common EBL spectral models: Franceschini-08 \cite{Franceschini:2008tp}, Finke-10 \cite{Finke:2009xi}, Gilmore-12 \cite{Gilmore:2011ks}, Franceschini-17 \cite{Franceschini:2017iwq}, and Saldana-Lopez-21 \cite{Saldana-Lopez:2020qzx}.
We can observe that these spectra exhibit similar distributions in the near-infrared regions, but demonstrate distinctly different distributions in the far-infrared regions, especially for the model Finke-10.
The energy spectra of the models Gilmore-12 and Franceschini-17 are relatively close overall, with only minor differences in the near-infrared regions.
Note that Franceschini-17 is an updated version of Franceschini-08, and compared to the latter, the photon density in the mid-infrared, far-infrared, and sub-millimeter regions of the former's energy spectrum is slightly lower.
Unlike the aforementioned models that rely on the Earth's observations, the latest model, Saldana-Lopez-21, adopts a completely new approach based solely on galaxy data to determine the evolving energy spectrum of the EBL, aiming to reduce the current uncertainties associated with high redshifts and infrared intensities.
Therefore, this model is more reliable as it can effectively reduce the undesired zodiacal light effect.
Additionally, it is noteworthy that there is another spectral model, CIBER-17 \cite{Matsuura:2017lub}, also derived from space observations. 
However, due to its worse statistical fit and the resulting unrealistic EBL, this model is not considered in our analysis.
For our purpose, here the EBL spectral model should be chosen as a relatively up-to-date one, and there should be certain distinctions among the energy spectra of these models.
Therefore, we select these three EBL spectral models in this work: Finke-10, Franceschini-17, and Saldana-Lopez-21, corresponding to the yellow, blue, and purple solid lines in figure~\ref{fig_ebl_model}, respectively.

\begin{table}[b]
\centering
\caption{BL Lac blazars with certain redshifts, sourced from \url{http://tevcat2.uchicago.edu}.}
\begin{ruledtabular}
\begin{tabular}{lclclc}
Source & redshift & Source & redshift  & Source & redshift \\
\hline
Markarian\,421   &   0.031    &  B2\,1811+31   &  0.117    &  1ES\,0347-121   & 0.188    \\                 
\fbox{Markarian\,501}   &   0.034    &  B3\,2247+381   &  0.1187    &  RBS\,0413   &  0.19   \\ 
1ES\,2344+514   &   0.044    &  TeV\,J1958-301   & 0.119329     &  RBS\,0723   &  0.198   \\                  
Markarian\,180   &    0.045   &  RGB\,J0710+591   & 0.125   & MRC\,0910-208   &  0.19802   \\    
1ES\,1959+650   &   0.048    &  TXS\,1515-273   &  0.1284    &  1ES\,1011+496   &  0.212   \\                  
TXS\,0210+515  &   0.049    &  H\,1426+428   &   0.129   &  TeV\,J1224+246   &   0.218  \\    
TeV\,J1517-243         &    0.049   &   1ES\,1215+303  &  0.131    &  TeV\,J0238-312   &  0.232   \\                  
1ES\,2037+521   &  0.053     &  TeV\,J1136+676   & 0.1342     &  \fbox{PKS\,0301-243}   &  0.2657   \\    
1ES\,1727+502   &  0.055     &  1ES\,0806+524   &  0.138    &  1ES\,0414+009   & 0.287    \\                  
PGC\,2402248  &   0.065    &  PKS\,1440-389   &  0.1385    &  OJ\,287   &  0.3056   \\    
PKS\,0548-322   &  0.069     &  \fbox{1ES\,0229+200}   &  0.1396    &  OT\,081   &  0.322   \\                  
BL\,Lacertae   &   0.069    &  TeV\,J1010-313   & 0.142639     &  S3\,1227+25   &  0.325   \\    
PKS\,2005-489   &  0.071     &  H\,2356-309   &   0.165   &  1ES\,0502+675   &  0.34   \\                  
RGB\,J0152+017   &  0.08     &  TeV\,J0812+026   &  0.1721    &  3C\,66A   &   0.34  \\    
1ES\,1741+196   &   0.084    &  1ES\,2322-409   &  0.1736    &  \fbox{PKS\,0447-439}   &  0.343   \\                  
TeV\,J0013-188   &  0.095     &  TeV\,J2001+438   &  0.1739    & 1ES\,0033+595    &  0.467   \\     
TeV\,J1221+282   &   0.102    &  TeV\,J0648+152   &   0.179   &   PKS\,0903-57  &  0.695   \\                   
1ES\,1312-423   &  0.105     &  1ES\,1218+304   &  0.182    & &  \\    
PKS\,2155-304  &  0.116     &  1ES\,1101-232   &   0.186   & &  \\                                         
\end{tabular}
\end{ruledtabular}
\label{tab_2}
\end{table} 

\subsection{The gamma-ray sources and data}

As mentioned before, here we should select the gamma-ray source with a relatively certain redshift. 
Additionally, we should choose various sources for analysis, ensuring that their redshifts range from low to high.
We list all currently known BL Lac blazars with certain redshifts in table~\ref{tab_2}, and their redshifts range from 0.031 to 0.695.
Among these sources, we select three sources, namely Markarian\,501 ($z_0=0.034$), 1ES\,0229+200 ($z_0=0.1396$), and PKS\,0301-243 ($z_0=0.2657$), with varying redshifts. 
Furthermore, we find that the two sources with the highest redshifts, 1ES\,0033+595 ($z_0=0.467$) and PKS\,0903-57 ($z_0=0.695$), do not have sufficient VHE gamma-ray observation data for ALP analysis, so we chose another source with a higher redshift, PKS\,0447-439 ($z_0=0.343$).
Therefore, in this work we select these four BL Lac blazars: Markarian\,501, 1ES\,0229+200, PKS\,0301-243, and PKS\,0447-439 for analysis.

Markarian\,501 ($\rm R.A.=16^h53^m52.2^s$, $\rm Dec.=+39^\circ45'37''$, J2000) is one of our closest and brightest BL Lac blazars, also known as Mrk\,501 and TeV\,J1653+397.
It was first discovered with the VHE emission greater than $300\rm \, GeV$ by the Whipple Observatory Gamma Ray Collaboration in 1996 \cite{Quinn:1996dj}.
Here the used VHE gamma-ray data of Markarian\,501 was recently measured by Fermi Large Area Telescope (Fermi-LAT) and High Altitude Water Cherenkov (HAWC) Gamma-Ray Observatory from June 2015 to July 2018 \cite{HAWC:2021obx}.

1ES\,0229+200 ($\rm R.A.=02^h32^m53.2^s$, $\rm Dec.=+20^\circ16'21''$, J2000) is known for emitting VHE gamma-rays in the hard TeV range, also known as TeV\,J0232+202.
It was first discovered in the Einstein Imaging Proportional Counter (IPC) Slew Survey in 1992 \cite{1992ApJS...80..257E} and was later categorized as a BL Lac object in 1993 \cite{1993ApJ...412..541S}.
In this work, the used VHE gamma-ray observation data of 1ES\,0229+200 was recently measured by Fermi-LAT and Major Atmospheric Gamma-ray Imaging Cherenkov (MAGIC) telescopes from September 2013 to December 2017 \cite{MAGIC:2022piy}.

PKS\,0301-243 ($\rm R.A.=03^h03^m23.49^s$, $\rm Dec.=-24^\circ07'35.86''$, J2000), also known as TeV\,J0303-241.
It was first identified as a blazar in 1988 \cite{1988ApJ...333..666I} and was later classified as a BL Lac object in 1996 \cite{1996A&A...311..384L}.
Here the used VHE gamma-ray data of PKS\,0301-243 was measured by Fermi-LAT and High Energy Stereoscopic System (H.E.S.S.) between September 2009 and December 2011 \cite{HESS:2013uta}.

PKS\,0447-439 ($\rm R.A.=04^h49^m28.2^s$, $\rm Dec.=-43^\circ50'12''$, J2000) is one of the brightest extragalactic objects listed in the Fermi Bright Source List and exhibits a hard spectrum within the MeV to GeV range, also known as TeV\,J0449-438.
It was first discovered in the radio band in 1981 \cite{1981MNRAS.194..693L} and was later identified as a bright BL Lac object \cite{Perlman:1998pk}.
Here the used VHE gamma-ray observations data of PKS\,0447-439 was measured by Fermi-LAT and H.E.S.S. from November 2009 to January 2010 \cite{HESS:2013mfa}. 

See figure~\ref{fig_dnde_ebl_null} for the experimental gamma-ray data of these sources with the blue and red points, respectively.

\subsection{The gamma-ray SEDs}

Here we show the gamma-ray SEDs under the null hypothesis. 
The VHE gamma-ray intrinsic spectrum $\Phi_{\rm int}(E)$ is selected as the power law with a super-exponential cut-off (SEPWL) model, which can be described by
\begin{eqnarray}
\Phi_{\rm int}(E)=N_0\left(E/E_0\right)^{-\Gamma}\exp\left(-\left(E/E_c\right)^d\right)\, , 
\end{eqnarray}
where $N_0$ is the normalization constant, $\Gamma$ is the spectral index, $E_c$ and $d$ are free parameters, and we fix $E_0$ with a typical value $1\, \rm GeV$. 
Then the chi-square value under the null hypothesis is given by 
\begin{eqnarray}
\chi_{\rm null}^2 = \sum_{i=1}^{N} \left(\dfrac{e^{-\tau}\Phi_{\rm int}(E_i) - \psi(E_i)}{\delta(E_i)}\right)^2\, ,
\end{eqnarray}
where $N$ is the gamma-ray spectral point number, $\psi$ and $\delta$ are the detected flux and its uncertainty, respectively.
For the experimental data of Markarian\,501, 1ES\,0229+200, PKS\,0301-243, and PKS\,0447-439, we have $N=33$, 13, 8, and 13, respectively.

\begin{figure*}[t]
\centering
\includegraphics[width=0.50\textwidth]{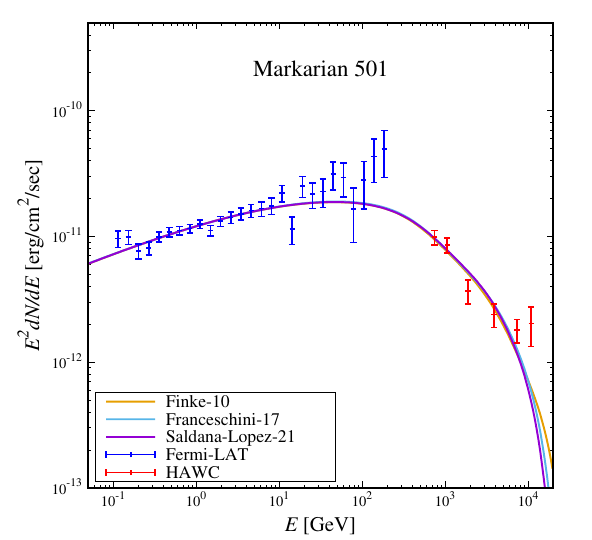}\includegraphics[width=0.50\textwidth]{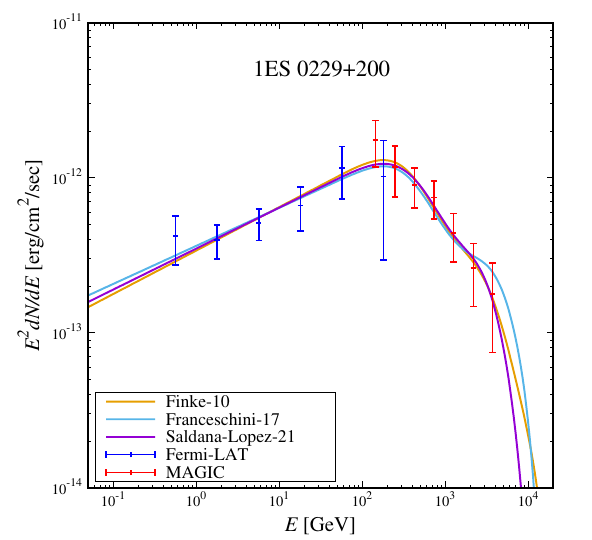}
\includegraphics[width=0.50\textwidth]{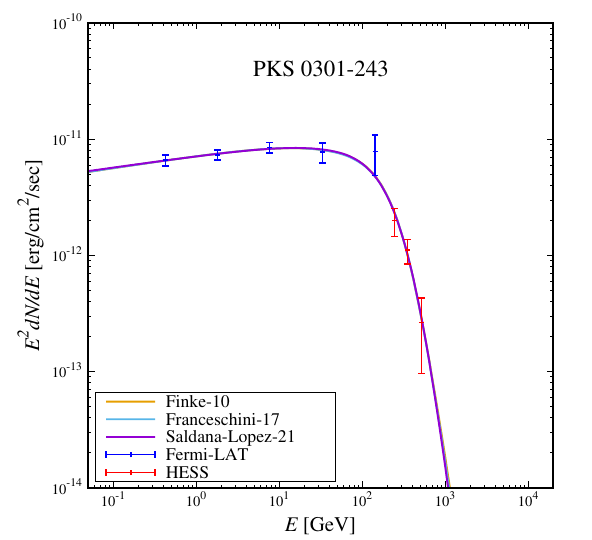}\includegraphics[width=0.50\textwidth]{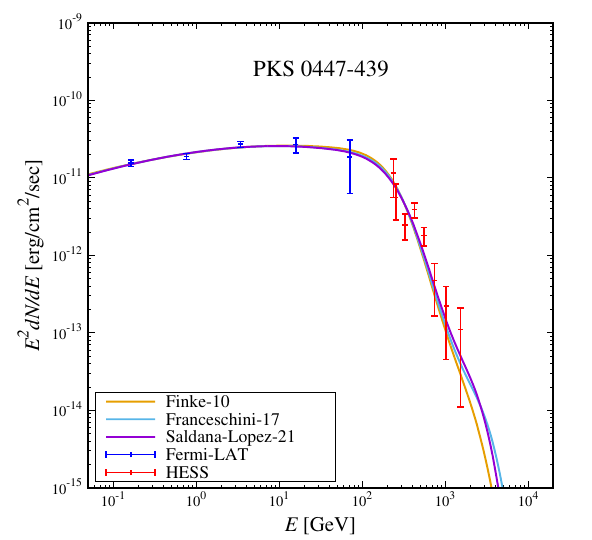}
\caption{The best-fit null hypothesis gamma-ray SEDs of Markarian\,501, 1ES\,0229+200, PKS\,0301-243, and PKS\,0447-439.
The yellow, blue, and purple lines correspond to the SEDs with the EBL spectral models Finke-10, Franceschini-17, and Saldana-Lopez-21, respectively.
The blue and red points represent the experimental data.}
\label{fig_dnde_ebl_null}
\end{figure*}

Using the above three EBL spectral models, we show the best-fit gamma-ray SEDs of Markarian\,501, 1ES\,0229+200, PKS\,0301-243, and PKS\,0447-439 under the null hypothesis in figure~\ref{fig_dnde_ebl_null}. 
The yellow, blue, and purple lines correspond to the null hypothesis SEDs with the EBL Finke-10, Franceschini-17, and Saldana-Lopez-21, respectively.
The best-fit chi-square values obtained with the different EBL spectral models are also listed in table~\ref{tab_3}, and their reduced chi-squares are all less than 2.
We find that for the low-redshift source Markarian\,501, their distributions are basically the same, except in the high-energy $\sim \mathcal{O}(10)\, \rm TeV$ region.
This is quite understandable, as the attenuation factor is greater at these energies.
This effect is more pronounced for the higher-redshift sources, 1ES\,0229+200 and PKS\,0447-439.
However, we find that for PKS\,0301-243, the SEDs do not vary significantly among different EBL models. 
This is mainly because there are relatively few observation points available for this source, and these observation points can be well fitted using different EBL models.
In addition, there is a scarcity of high-energy data for this source.
Overall, we find that the impact of different EBL spectral models on the SEDs of these sources under the null hypothesis is relatively small, with the exception of the high-energy range.

\begin{table}[b]
\centering
\caption{The best-fit null hypothesis chi-square $\chi^2_{\rm null}$ values of Markarian\,501, 1ES\,0229+200, PKS\,0301-243, and PKS\,0447-439 with the EBL spectral models Finke-10, Franceschini-17, and Saldana-Lopez-21.}
\begin{ruledtabular}
\begin{tabular}{llcc}
Source & EBL model & $\chi^2_{\rm null}$ & $\chi^2_{\rm null}/{\rm d.o.f.}$  \\
\hline
Markarian\,501  &  Finke-10                   &  41.08      &   1.42    \\
(33 points)         &  Franceschini-17        &  41.09     &   1.42    \\
                          &  Saldana-Lopez-21    &  43.07     &   1.49    \\
\hline                       
1ES\,0229+200  &  Finke-10                   &   2.54     &    0.28    \\
(13 points)          &  Franceschini-17        &   3.27    &    0.36   \\
                           &  Saldana-Lopez-21    &   2.47    &    0.27    \\
\hline                   
PKS\,0301-243  &  Finke-10                    &   1.77     &   0.44     \\
(8 points)            &  Franceschini-17        &    1.78    &   0.45    \\
                           &  Saldana-Lopez-21    &    1.74    &   0.44     \\
\hline                   
PKS\,0447-439  &  Finke-10                   &   17.16    &   1.91     \\
(13 points)         &  Franceschini-17        &   16.18    &   1.80    \\
                          &  Saldana-Lopez-21    &   14.36    &    1.60    \\
\end{tabular}
\end{ruledtabular}
\label{tab_3}
\end{table}  

\section{Photon-ALP conversions in astrophysical magnetic fields}
\label{sec_ALP}

In this section, we discuss the photon-ALP conversions in astrophysical magnetic fields.
We first introduce the photon-ALP conversions in the inhomogeneous magnetic field, then we discuss the setup of astrophysical magnetic field parameters.

\subsection{Photon-ALP conversions in the magnetic field}

Before discussing the photon-ALP conversions in the inhomogeneous astrophysical magnetic field, we have already provided the general conversions in the homogeneous magnetic field in appendix~\ref{appendix_1}.

In the real astrophysical environment, the magnetic field is inhomogeneous and can be random.
In order to obtain the photon-ALP conversion probability in the inhomogeneous magnetic field, the magnetic field is usually simulated with the domain-like structure and each domain can be regarded as homogeneous.
In this case, the photon-ALP system can be described by the density matrix
\begin{eqnarray}
\rho(x_3) = 
\left(
\begin{array}{c}
A_1(x_3)\\
A_2(x_3)\\
a(x_3)
\end{array}
\right)\otimes
\left(
\begin{array}{c}
A_1(x_3), A_2(x_3), a(x_3)
\end{array}
\right)^* \, ,
\end{eqnarray} 
which satisfies the Von Neumann-like commutator equation
\begin{eqnarray}
i\dfrac{{\rm d}\rho(x_3)}{{\rm d}x_3}= \rho(x_3)\mathcal{M}^\dagger(E,x_3,\theta)-\mathcal{M}(E,x_3,\theta)\rho(x_3) \, ,
\label{Neumann-like_equation}
\end{eqnarray} 
where $x_3$ is the direction of propagation, $A_1$ and $A_2$ are the linear polarization amplitudes of the photon in the perpendicular directions ($x_1$, $x_2$), and $a$ is the ALP.
Notice that in a case where the transversal magnetic field $B_T$ is not aligned along the direction of $x_2$ and forms an angle $\theta$, the mixing matrix $\mathcal{M}(E,x_3)$ should be
\begin{eqnarray}
\mathcal{M}(E,x_3)\to \mathcal{V}^\dag(\theta)\mathcal{M}(E,x_3)\mathcal{V}(\theta)\, ,
\end{eqnarray}
with
\begin{eqnarray}
\mathcal{V}(\theta)=
\left(
\begin{array}{ccc}
\cos \theta~ &  -\sin \theta~  & 0  \\
\sin \theta~ &  \cos \theta~  & 0  \\
0 &  0  & 1  \\
\end{array}
\right)\, ,
\end{eqnarray}
and we have the mixing matrix
\begin{eqnarray}
\mathcal{M}(E,x_3,\theta)=
\left(
\begin{array}{ccc}
\Delta_{11}(E,x_3)  &  0 & \Delta_{a \gamma}(x_3)\sin \theta \\
0  &  \Delta_{22}(E,x_3) & \Delta_{a \gamma}(x_3)\cos \theta \\
\Delta_{a \gamma}(x_3)\sin \theta~  &  \Delta_{a \gamma}(x_3)\cos \theta~ & \Delta_{aa}(E)
\end{array}
\right)\, .
\label{mixing_matrix}
\end{eqnarray}
The $\Delta$ terms in eq.~(\ref{mixing_matrix}) can be found in appendix~\ref{appendix_1} within the homogeneous magnetic field.
The solution of eq.~(\ref{Neumann-like_equation}) can be described by
\begin{eqnarray}
\rho(x_3)=\mathcal{T}(E,x_3,\theta)\rho(0)\mathcal{T}^\dagger(E,x_3,\theta) \, ,
\end{eqnarray} 
where $\mathcal{T}(E,x_3,\theta)$ is the whole transport matrix of the $n$ domains
\begin{eqnarray}
\mathcal{T}(E,x_3,\theta)= \prod^{n}_{i=1}\mathcal{T}(E_i,x_{3,i},\theta_i) \, ,
\end{eqnarray} 
and $\rho(0)$ is the initial density matrix
\begin{eqnarray}
\rho(0)=\dfrac{1}{2}
\left(
\begin{array}{ccc}
1~ & 0~ & 0 \\
0~ & 1~ & 0 \\
0~ & 0~ & 0 
\end{array}
\right)\, .
\end{eqnarray}
Then the photon-ALP-photon conversion probability, or the final photon survival probability, can be described by
\begin{eqnarray}
\mathcal{P}_{\gamma\gamma} = {\rm Tr}\left[\left(\rho_{11}+\rho_{22}\right)\mathcal{T}(E,x_3,\theta)\rho(0)\mathcal{T}^\dagger(E,x_3,\theta) \right] \, ,
\end{eqnarray} 
with the matrices
\begin{eqnarray}
{\rho}_{11} = 
\left(\begin{array}{ccc}
1~ & 0~ & 0 \\
0~ & 0~ & 0 \\
0~ & 0~ & 0 
\end{array}\right)\, , \quad
{\rho}_{22} = 
\left(\begin{array}{ccc}
0~ & 0~ & 0 \\
0~ & 1~ & 0 \\
0~ & 0~ & 0 
\end{array}\right)\, .
\end{eqnarray}
See also ref.~\cite{DeAngelis:2011id} for more details in the calculations.
Note that specific analysis is required for the specific magnetic field models. 
Finally, we can obtain a set of the final photon survival probability $\mathcal{P}_{\gamma\gamma}$.

\subsection{Astrophysical magnetic fields setup}

Here we discuss the astrophysical magnetic fields setup associated the photon-ALP beam propagating from the VHE gamma-ray source region to the Earth.
Generally, this process is composed of three parts: (1) the source region, (2) the extragalactic space, and (3) the Milky Way.
 
Firstly, for (1) the source region of the BL Lac object, the blazar jet magnetic field can be described by the poloidal and toroidal components.
We consider the photon-ALP conversions in the transverse magnetic field model $B(r) = B_0(r/r_{\rm VHE})^{-1}$ with the electron density model $n_{\rm el}(r) = n_0(r/r_{\rm VHE})^{-2}$, where $r_{\rm VHE}\sim R_{\rm VHE}/\theta_{\rm jet}$ represents the distance between the VHE emission region and the central black hole of the source, $R_{\rm VHE}$ represents the radius of the VHE emission, $\theta_{\rm jet}$ represents the angle between the jet axis and the line of sight, $B_0$ and $n_0$ represent the core magnetic field and electron density at the distance $r_{\rm VHE}$, respectively.
Since $n_0$ has a minimal impact on the final result, we take $n_0=1\times 10^3\, \rm cm^{-3}$ as a typical value in this work.
Here we also consider the energy transformation with the Doppler factor, $\delta_{\rm D}=E_L/E_j$, where $E_L$ and $E_j$ represent the energy in the laboratory and co-moving frames, respectively.
For the jet region $r > 1\rm\, kpc$, the magnetic field is taken as zero.
Additionally, for the host galaxy region of the source, the photon-ALP conversion effect can be totally neglected.
Then for (2) the extragalactic space, we just need to consider the EBL absorption effect on the VHE gamma-rays due to the electron-positron pair-production process, see also section~\ref{EBL_absorption}.
Since the strength of magnetic field in the extragalactic space is small with the upper limit $\sim\mathcal{O}(1)\, \rm nG$ \cite{Ade:2015cva, Pshirkov:2015tua}, the photon-ALP conversion effect will be weak, and thus we do not consider the photon-ALP conversions in this part.
Finally, in (3) the Milky Way, we should consider the photon-ALP conversions again in the Galactic magnetic field.
Generally, this magnetic field can be modeled with the disk and halo components (parallel to the Galactic plane), and the so-called ``X-field" component (out-of-plane) at the Galactic center \cite{Jansson:2012pc, Jansson:2012rt}.
See also refs.~\cite{Planck:2016gdp, Unger:2023lob} for the latest version of this Galactic magnetic field model.

Here we list the blazar jet magnetic field model parameters $B_0$, $R_{\rm VHE}$, $\theta_{\rm jet}$, $r_{\rm VHE}$, and $\delta_{\rm D}$ of Markarian\,501, 1ES\,0229+200, PKS\,0301-243, and PKS\,0447-439 in table~\ref{tab_4}. 
On the one hand, some magnetic field parameters are not directly provided; on the other hand, since our main goal here is not to impose specific constraints on the ALP parameter space, we have assigned typical values to some of these parameters.
As we use the same magnetic field model parameters for each EBL model, our main conclusions will not be affected.
 
\begin{table}[b]
\centering
\caption{The blazar jet magnetic field model parameters of Markarian\,501, 1ES\,0229+200, PKS\,0301-243, and PKS\,0447-439.}
\begin{ruledtabular}
\begin{tabular}{lcccccc}
Source & $B_0$ (mG) & $R_{\rm VHE}$ ($10^{17}\, \rm cm$)  &  $\theta_{\rm jet}$ (deg) & $r_{\rm VHE}$ ($10^{17}\, \rm cm$) &  $\delta_{\rm D}$ & ref. \\
\hline 
Markarian\,501   & 20           & 1.0             &  3.0          &  19.1  &  13           &  \cite{HAWC:2021obx} \\
1ES\,0229+200  & $\sim2$   &  $\sim0.2$ &  2.5          &  4.6    & $\sim50$  & \cite{Aliu:2013pya} \\
PKS\,0301-243  & 20            &  1.3            &  $\sim2$  &  37.2  & 27             & \cite{HESS:2013uta} \\
PKS\,0447-439  & 20            & 0.65           &  $\sim2$  &  18.6  & $\sim51$  &  \cite{HESS:2013mfa} \\
\end{tabular}
\end{ruledtabular}
\label{tab_4}
\end{table} 

\begin{figure*}[!htbp]
\centering
\subfigcapskip=-3.45pt
\subfigbottomskip=0pt
\subfigure[Markarian\,501 with Finke-10.]{\includegraphics[width=8.3cm]{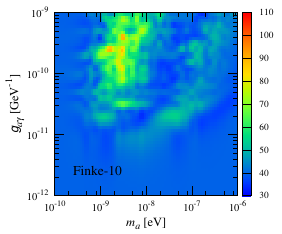}}\subfigure[1ES\,0229+200 with Finke-10.]{\includegraphics[width=8.3cm]{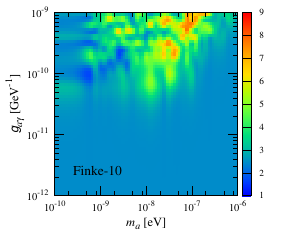}}
\subfigure[Markarian\,501 with Franceschini-17.]{\includegraphics[width=8.3cm]{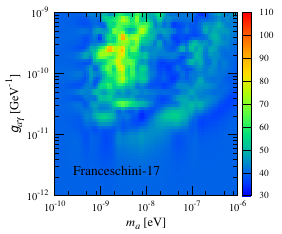}}\subfigure[1ES\,0229+200 with Franceschini-17.]{\includegraphics[width=8.3cm]{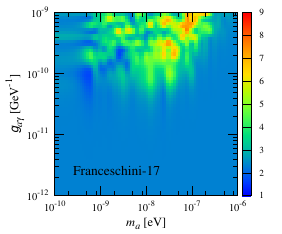}}
\subfigure[Markarian\,501 with Saldana-Lopez-21.]{\includegraphics[width=8.3cm]{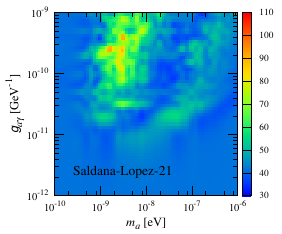}}\subfigure[1ES\,0229+200 with Saldana-Lopez-21.]{\includegraphics[width=8.3cm]{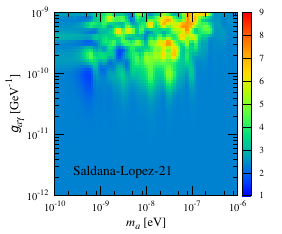}}
\caption{The best-fit ALP assumption chi-square $\chi_{\rm ALP}^2$ distributions.
The left and right panels correspond to Markarian\,501 and 1ES\,0229+200, respectively.
The top, middle, and bottom panels correspond to the EBL models Finke-10, Franceschini-17, and Saldana-Lopez-21, respectively.}
\label{fig_contour_ebl_501_200}
\end{figure*}

\begin{figure*}[!htbp]
\centering
\subfigcapskip=-3.45pt
\subfigbottomskip=0pt
\subfigure[PKS\,0301-243 with Finke-10.]{\includegraphics[width=8.3cm]{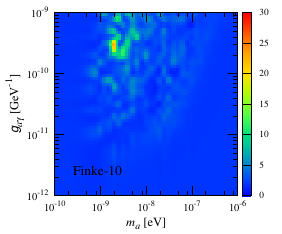}}\subfigure[PKS\,0447-439 with Finke-10.]{\includegraphics[width=8.3cm]{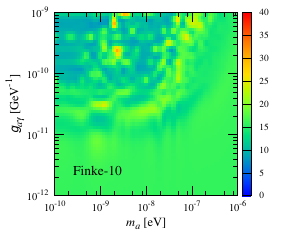}}
\subfigure[PKS\,0301-243 with Franceschini-17.]{\includegraphics[width=8.3cm]{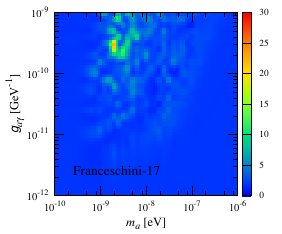}}\subfigure[PKS\,0447-439 with Franceschini-17.]{\includegraphics[width=8.3cm]{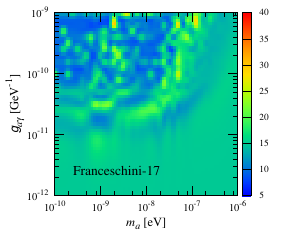}}
\subfigure[PKS\,0301-243 with Saldana-Lopez-21.]{\includegraphics[width=8.3cm]{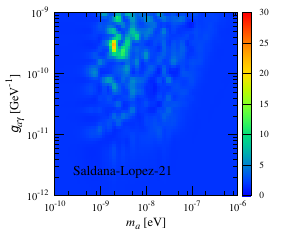}}\subfigure[PKS\,0447-439 with Saldana-Lopez-21.]{\includegraphics[width=8.3cm]{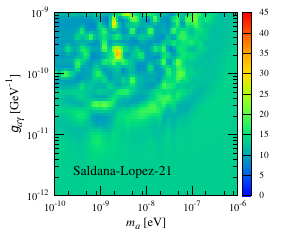}}
\caption{The best-fit ALP assumption chi-square $\chi_{\rm ALP}^2$ distributions.
The left and right panels correspond to PKS\,0301-243 and PKS\,0447-439, respectively.
The top, middle, and bottom panels correspond to the EBL models Finke-10, Franceschini-17, and Saldana-Lopez-21, respectively.}
\label{fig_contour_ebl_243_439}
\end{figure*}

\begin{figure*}[t]
\centering
\includegraphics[width=0.50\textwidth]{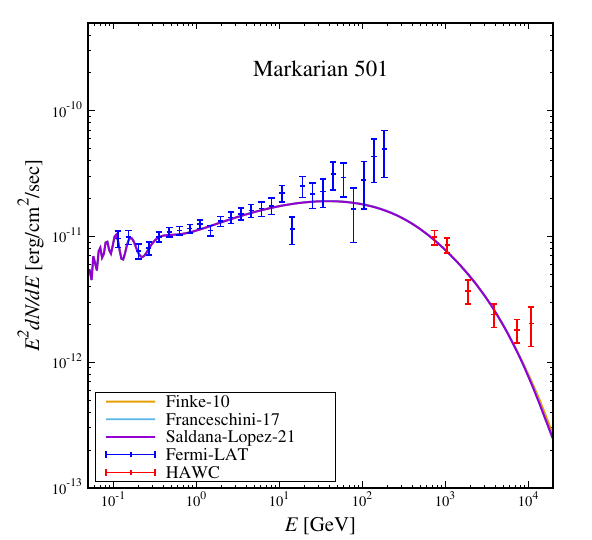}\includegraphics[width=0.50\textwidth]{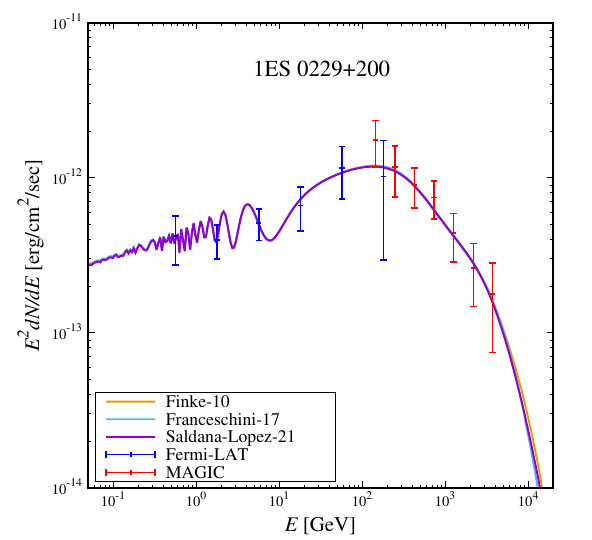}
\includegraphics[width=0.50\textwidth]{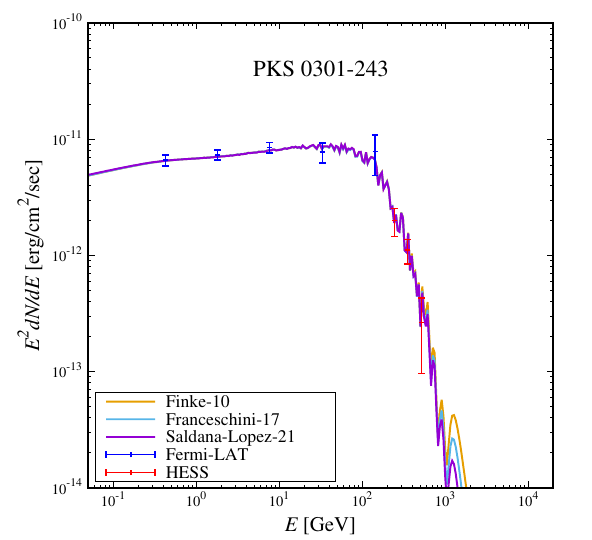}\includegraphics[width=0.50\textwidth]{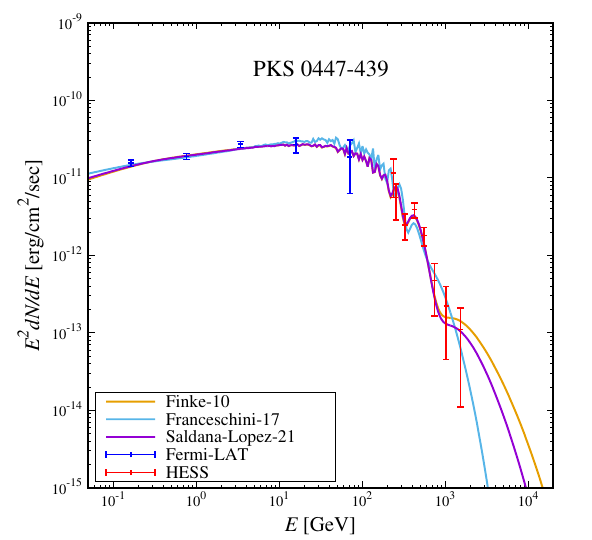}
\caption{The best-fit ALP assumption gamma-ray SEDs corresponding to $\chi^2_{\rm min}$ of Markarian\,501, 1ES\,0229+200, PKS\,0301-243, and PKS\,0447-439.
The yellow, blue, and purple lines correspond to the SEDs with the EBL spectral models Finke-10, Franceschini-17, and Saldana-Lopez-21, respectively.
The blue and red points represent the experimental data.}
\label{fig_dnde_ebl_alp_min}
\end{figure*}

\section{Analysis and results}
\label{sec_results}

In this section, we show our analysis and results under the ALP assumption with the different EBL spectral models.
After considering the above photon-ALP conversion effect, we can derive the final photon survival probability $\mathcal{P}_{\gamma\gamma}$, then the chi-square value under the ALP assumption is given by 
\begin{eqnarray}
\chi_{\rm ALP}^2 = \sum_{i=1}^{N} \left(\dfrac{\mathcal{P}_{\gamma\gamma}\Phi_{\rm int}(E_i) - \psi(E_i)}{\delta(E_i)}\right)^2\, .
\end{eqnarray}
Notice that in the calculations the EBL absorption effect in the extragalactic space is also included in the final photon survival probability.
For one ALP $\{m_a, g_{a\gamma}\}$ parameter set, we can derive one best-fit $\chi^2_{\rm ALP}$ under the ALP assumption, and also the best-fit $\chi^2_{\rm ALP}$ distribution in the whole ALP parameter plane.
By utilizing this chi-square distribution, we can obtain the $\Delta\chi^2_{\rm ALP}$ at the particular confidence level to establish the corresponding ALP bound. 
However, in this work, we do not intend to set any limits on ALP, but rather focus on the EBL absorption effect using different EBL spectral models.

\begin{table}[b]
\centering
\caption{The minimum best-fit ALP assumption chi-square $\chi^2_{\rm min}$ values of Markarian\,501, 1ES\,0229+200, PKS\,0301-243, and PKS\,0447-439 with the EBL spectral models Finke-10, Franceschini-17, and Saldana-Lopez-21.}
\begin{ruledtabular}
\begin{tabular}{lcccc}
Source & EBL model & $\chi^2_{\rm min}$ &   $m_a$ (${\rm eV}$) & $g_{a\gamma}$ ($\rm GeV^{-1}$) \\ 
\hline
Markarian\,501  &  Finke-10                   &    30.52     &   $1.0\times 10^{-9}$    &   $4.0\times 10^{-10}$   \\ 
(33 points)         &  Franceschini-17        &   30.66     &   $1.0\times 10^{-9}$    &   $4.0\times 10^{-10}$   \\
                          &  Saldana-Lopez-21    &   30.90    &   $1.0\times 10^{-9}$    &   $4.0\times 10^{-10}$   \\
\hline                       
1ES\,0229+200  &  Finke-10                   &   1.46      &    $2.5\times 10^{-9}$    &   $2.0\times 10^{-10}$   \\
(13 points)          &  Franceschini-17        &   1.46     &    $2.5\times 10^{-9}$    &   $2.0\times 10^{-10}$   \\
                           &  Saldana-Lopez-21    &    1.50    &    $2.5\times 10^{-9}$    &   $2.0\times 10^{-10}$   \\
\hline                   
PKS\,0301-243  &  Finke-10                    &    0.48     &   $1.0\times 10^{-7}$    &   $6.3\times 10^{-10}$   \\
(8 points)            &  Franceschini-17        &    0.48     &   $1.0\times 10^{-7}$    &   $6.3\times 10^{-10}$   \\
                           &  Saldana-Lopez-21    &    0.52    &   $1.0\times 10^{-7}$    &   $6.3\times 10^{-10}$   \\
\hline                   
PKS\,0447-439  &  Finke-10                   &     4.79     &    $3.2\times 10^{-8}$    &   $1.6\times 10^{-10}$   \\
(13 points)         &  Franceschini-17        &     5.92    &     $4.0\times 10^{-8}$    &   $4.0\times 10^{-10}$   \\
                          &  Saldana-Lopez-21    &     4.64    &     $3.2\times 10^{-8}$    &   $1.6\times 10^{-10}$   \\
\end{tabular}
\end{ruledtabular}
\label{tab_5}
\end{table} 

We first show in figures~\ref{fig_contour_ebl_501_200} and \ref{fig_contour_ebl_243_439} the best-fit chi-square $\chi_{\rm ALP}^2$ distributions of Markarian\,501, 1ES\,0229+200, PKS\,0301-243, and PKS\,0447-439 under the ALP assumption in the $\{m_a, g_{a\gamma}\}$ plane.
The three EBL spectral models Finke-10, Franceschini-17, and Saldana-Lopez-21 are arranged from top to bottom.
When comparing the panels for the same source, we find that there are no significant changes in the $\chi_{\rm ALP}^2$ values for the same parameter set across different EBL spectral models.
Here the minimum best-fit chi-square in the $\{m_a, g_{a\gamma}\}$ plane can be defined as $\chi^2_{\rm min}$, which are listed in table~\ref{tab_5}.
For the sources Markarian\,501, 1ES\,0229+200, and PKS\,0301-243, we find that the ALP parameter points corresponding to $\chi^2_{\rm min}$ from the three different EBL models all coincide at the same point. 
However, for PKS\,0447-439, only two of the EBL models have their minimum chi-square values corresponding to the same location.
Next we show in figure~\ref{fig_dnde_ebl_alp_min} the gamma-ray SEDs corresponding to $\chi^2_{\rm min}$.
We find that for the sources Markarian\,501 and 1ES\,0229+200, the SEDs corresponding to different EBL models are almost identical. 
However, for the sources PKS\,0301-243 and PKS\,0447-439, there are certain differences in their distributions at high-energy ranges, particularly for PKS\,0447-439. 
The noticeable variations in the SEDs of PKS\,0447-439 can be partly attributed to the different magnetic field configuration (resulting from different ALP parameter points) corresponding to the EBL model Franceschini-17 and, additionally, to the high redshift of PKS\,0447-439.

\begin{figure*}[t]
\centering
\subfigcapskip=-3.1pt
\subfigbottomskip=0pt
\subfigure[Markarian\,501.]{\includegraphics[width=8.3cm]{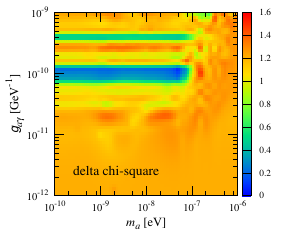}}\subfigure[1ES\,0229+200.]{\includegraphics[width=8.3cm]{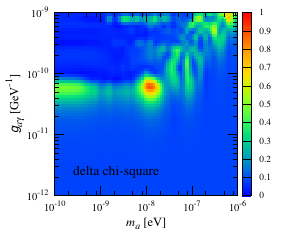}}
\subfigure[PKS\,0301-243.]{\includegraphics[width=8.3cm]{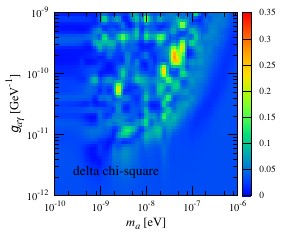}}\subfigure[PKS\,0447-439.]{\includegraphics[width=8.3cm]{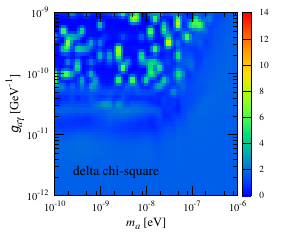}}
\caption{The delta chi-square $\chi_d^2$ distributions of Markarian\,501, 1ES\,0229+200, PKS\,0301-243, and PKS\,0447-439.}
\label{fig_contour_ebl_alp_com}
\end{figure*}

\begin{figure*}[t]
\centering
\includegraphics[width=0.50\textwidth]{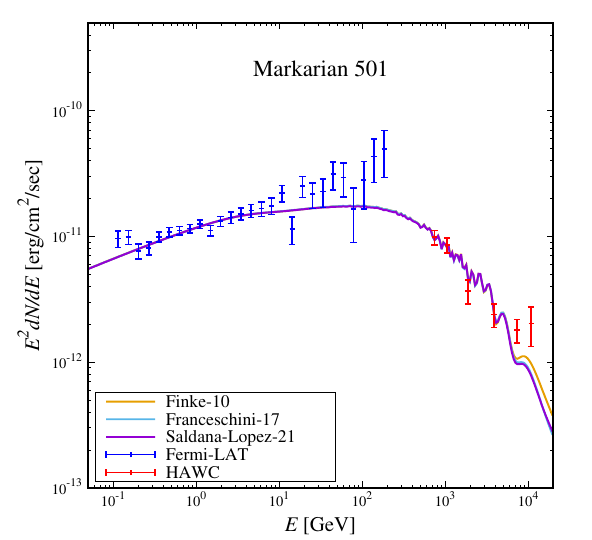}\includegraphics[width=0.50\textwidth]{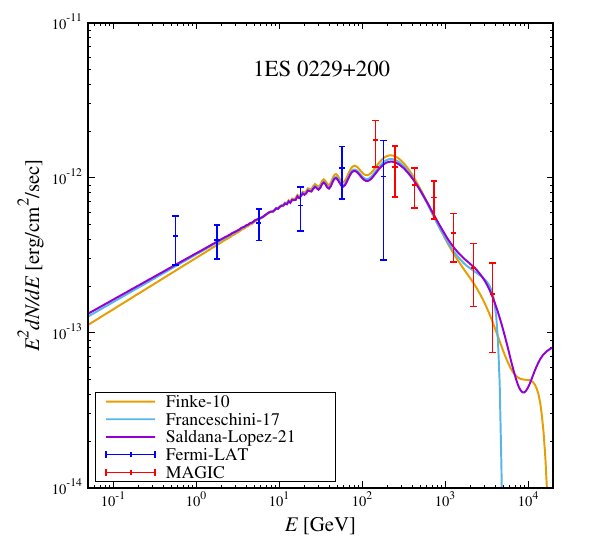}
\includegraphics[width=0.50\textwidth]{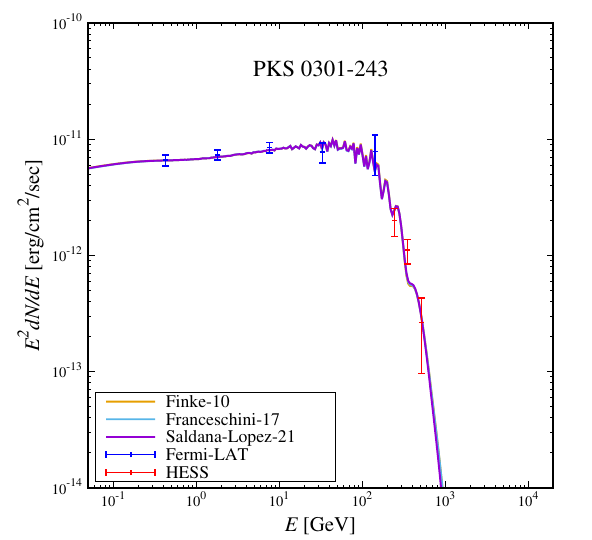}\includegraphics[width=0.50\textwidth]{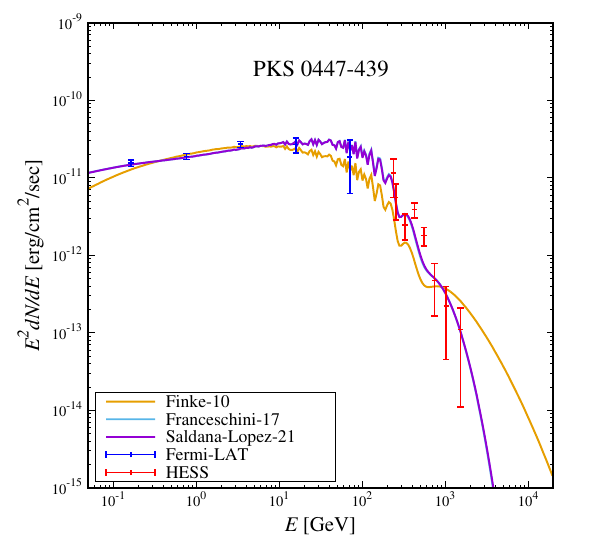}
\caption{The best-fit ALP assumption gamma-ray SEDs corresponding to $\chi^2_{d, \rm max}$ of Markarian\,501, 1ES\,0229+200, PKS\,0301-243, and PKS\,0447-439.
The yellow, blue, and purple lines correspond to the SEDs with the EBL spectral models Finke-10, Franceschini-17, and Saldana-Lopez-21, respectively.
The blue and red points represent the experimental data.}
\label{fig_dnde_ebl_alp_max}
\end{figure*} 

Therefore, in order to characterize the difference in chi-square between the three different EBL spectral models in this work, it is necessary to define a new delta chi-square for each $\{m_a, g_{a\gamma}\}$ set
\begin{eqnarray}
\begin{aligned}
\chi_d^2&=\dfrac{1}{6}\sum_{i=1}^3 \sum_{j=1}^3 \left(1-\delta_{ij}\right) \Big|\chi^2_{{\rm ALP},i}-\chi^2_{{\rm ALP},j}\Big|\\
&= \dfrac{1}{3} \left(\Big|\chi^2_{{\rm ALP},1}-\chi^2_{{\rm ALP},2}\Big| + \Big|\chi^2_{{\rm ALP},1}-\chi^2_{{\rm ALP},3}\Big| + \Big|\chi^2_{{\rm ALP},2}-\chi^2_{{\rm ALP},3}\Big|\right)\, ,
\label{delta_chi-square}
\end{aligned}
\end{eqnarray}
where $i$ and $j$ represent the number of the three EBL models, and $\delta_{ij}$ is the Kronecker delta function
\begin{eqnarray}
\delta_{ij}=\begin{cases}
1\, , & i=j\\ 
0\, . & i\ne j 
\end{cases}
\end{eqnarray}
Notice that eq.~(\ref{delta_chi-square}) is only used to quantify the chi-square variation, and any subtle chi-square differences among the three EBL spectral models can be reflected by this value.
Figure~\ref{fig_contour_ebl_alp_com} shows the distributions of this delta chi-square in the $\{m_a, g_{a\gamma}\}$ plane of Markarian\,501, 1ES\,0229+200, PKS\,0301-243, and PKS\,0447-439, respectively.
From this plot, we can clearly observe the chi-square differences and their corresponding ALP parameter points.
The results indicate that across the entire parameter plane, the chi-square values of the sources Markarian\,501, 1ES\,0229+200, and PKS\,0301-243 are consistently small and remain below 2. 
Conversely, within the same parameter plane, PKS\,0447-439 demonstrates numerous chi-square values that are significantly larger.
Here the maximum delta chi-square in the $\{m_a, g_{a\gamma}\}$ plane can be defined as $\chi^2_{d, \rm max}$, which are listed in table~\ref{tab_6}.
Then we show in figure~\ref{fig_dnde_ebl_alp_max} the gamma-ray SEDs corresponding to $\chi^2_{d, \rm max}$.
We observe that for the sources Markarian\,501 and PKS\,0301-243, there is little difference in their SEDs across different EBL models. 
However, for 1ES\,0229+200 and PKS\,0447-439, under different EBL models, there are significant differences in the high-energy range. 
Note that for PKS\,0447-439, the spectra displayed here are all for the same ALP parameter point, which differs from the distributions shown in figure~\ref{fig_dnde_ebl_alp_min}.

\begin{table}[h]
\centering
\caption{The maximum delta chi-square $\chi^2_{d, \rm max}$ values of Markarian\,501, 1ES\,0229+200, PKS\,0301-243, and PKS\,0447-439.}
\begin{ruledtabular}
\begin{tabular}{lccc}
Source & $\chi^2_{d, \rm max}$ & $m_a$ (${\rm eV}$) & $g_{a\gamma}$ ($\rm GeV^{-1}$) \\
\hline
Markarian\,501   &   1.59      &  $1.6\times 10^{-7}$    &   $1.0\times 10^{-10}$   \\                  
1ES\,0229+200  &   0.95      &  $1.3\times 10^{-8}$    &  $6.3\times 10^{-11}$    \\                 
PKS\,0301-243  &    0.31      &  $4.0\times 10^{-8}$    &  $2.0\times 10^{-10}$    \\                  
PKS\,0447-439  &   13.00     &  $3.2\times 10^{-8}$  &  $3.2\times 10^{-10}$    \\
\end{tabular}
\end{ruledtabular}
\label{tab_6}
\end{table}

\section{Conclusion}
\label{sec_onclusion}

In summary, we have investigated the impact of the EBL absorption effect on photon-ALP conversions from the VHE gamma-ray spectral irregularities.
For our purpose, we select four BL Lac blazars: Markarian\,501 ($z_0=0.034$), 1ES\,0229+200 ($z_0=0.1396$), PKS\,0301-243 ($z_0=0.2657$), and PKS\,0447-439 ($z_0=0.343$).
These gamma-ray sources have a relatively certain redshift, ranging from low to high, making them a suitable choice for exploring the impact of the EBL absorption effect.
On the other hand, we select three EBL spectral models: Finke-10, Franceschini-17, and Saldana-Lopez-21.
These models are the latest ones available, and there are distinct differences among them.
More importantly, the observations for Saldana-Lopez-21 originate from space, providing greater reliability than those conducted from the Earth.

We first discuss the EBL absorption effect on the gamma-ray SEDs with these three EBL spectral models.
Then we consider the photon-ALP conversions in astrophysical magnetic fields.
This includes a discussion on photon-ALP conversions in both inhomogeneous magnetic fields and the astrophysical magnetic field configurations. 
The best-fit chi-square distributions of these EBL models under the ALP assumption in the ALP parameter $\{m_a, g_{a\gamma}\}$ plane are given (see figure~\ref{fig_dnde_ebl_null}), showing similar distributions.
From the gamma-ray SEDs corresponding to $\chi^2_{\rm min}$, we find that for Markarian\,501 and 1ES\,0229+200, the SEDs corresponding to different EBL models are almost identical. 
However, for PKS\,0301-243 and PKS\,0447-439, there are certain differences in their distributions at high-energy ranges.
For comparison, we define a new delta chi-square $\chi_d^2$ to quantify the chi-square difference.
The distributions of $\chi_d^2$ and the gamma-ray SEDs corresponding to the maximum delta chi-square $\chi^2_{d, \rm max}$ in the $\{m_a, g_{a\gamma}\}$ plane are also shown (see figures~\ref{fig_contour_ebl_501_200}, \ref{fig_contour_ebl_243_439}, and \ref{fig_contour_ebl_alp_com}).
Finally, we find that there is only a minor influence from the different EBL models at the low-redshift gamma-ray axionscope. 
However, as the redshift of the sources increases, this impact becomes more pronounced, which can be observed from the gamma-ray SEDs (see figures~\ref{fig_dnde_ebl_alp_min} and \ref{fig_dnde_ebl_alp_max}). 
On the other hand, the uncertainty in the observed spectra can also directly affect the results. 
Overall, for the investigation of photon-ALP coupling from the low-redshift gamma-ray sources, the selection of the EBL spectral model Saldana-Lopez-21 is sufficient, as it provides a reliable absorption model for this scenario.
Additionally, this absorption effect is expected to become more significant in the next generation of higher-energy gamma-ray observation experiments.

\medskip\noindent{\bf Acknowledgments.}---%
W.C. is supported by the National Key R\&D Program of China (Grant No.~2023YFA1607104), the National Natural Science Foundation of China (NSFC) (Grants No.~11775025 and No.~12175027), and the Fundamental Research Funds for the Central Universities (Grant No.~2017NT17).
Y.F.Z. is supported by the National Key R\&D Program of China (Grant No.~2017YFA0402204), the CAS Project for Young Scientists in Basic Research YSBR-006, and the NSFC (Grants No.~11821505, No.~11825506, and No.~12047503).

\appendix

\section{General photon-ALP conversions}
\label{appendix_1}
 
Here we present the general photon-ALP conversions in the homogeneous magnetic field and the calculation of the photon-ALP conversion probability \cite{Raffelt:1987im, DeAngelis:2011id, Li:2022mcf}. 
The photon-ALP system ($A_1$, $A_2$, and $a$) can be described by 
\begin{eqnarray}
\psi(x_3) = 
\left(\begin{array}{c}
A_1(x_3)\\
A_2(x_3)\\
a(x_3)
\end{array}\right)\, ,
\end{eqnarray}
where $x_3$ is the direction of propagation, $A_1$ and $A_2$ represent the linear polarization amplitudes of the photons in the perpendicular directions ($x_1$, $x_2$)
\begin{eqnarray}
|A_1\rangle=\left( \begin{array}{c} 1 \\ 0 \\ 0  \end{array}\right)\, , \quad |A_2\rangle = \left(\begin{array}{c} 0 \\ 1 \\ 0  \end{array}\right)\, , \quad |a\rangle=\left(\begin{array}{c}0 \\ 0 \\ 1  \end{array}\right)\, .
\end{eqnarray}
Then the equation of motion for the photon-ALP system in the magnetic field can be described by
\begin{eqnarray}
\left(i \dfrac{\rm d}{{\rm d} x_3}+E+\mathcal{M}(E,x_3)\right)\psi(x_3) =0\, ,
\end{eqnarray}
where $E$ is the photon-ALP beam energy, and $\mathcal{M}(E,x_3)$ is the mixing matrix 
\begin{eqnarray}
\mathcal{M}(E,x_3)=
\left(\begin{array}{ccc}
\Delta_{11}(E,x_3)~  &  \Delta_{12}(E,x_3)~ & \Delta_{a \gamma, 1}(x_3)\\
\Delta_{21}(E,x_3)~  &  \Delta_{22}(E,x_3)~ & \Delta_{a \gamma, 2}(x_3)\\
\Delta_{a \gamma, 1}(x_3)~  &  \Delta_{a \gamma, 2}(x_3)~  & \Delta_{aa}(E)
\end{array}\right)\, .
\end{eqnarray}
These terms are given by
\begin{eqnarray}
\Delta_{11}(E,x_3)&=&\Delta_{\rm pl}(E,x_3)+2\Delta_{\rm QED}(E,x_3)  + \Delta_{\rm CMB}(E)\, ,\\
\Delta_{22}(E,x_3)&=&\Delta_{\rm pl}(E,x_3)+\dfrac{7}{2} \Delta_{\rm QED}(E,x_3)  + \Delta_{\rm CMB}(E)\, ,
\end{eqnarray}
with
\begin{eqnarray}
\Delta_{\rm pl}(E,x_3)&=&-\dfrac{\omega_{\rm pl}^2(x_3)}{2E}\simeq -1.08 \times 10^{-1} \left(\dfrac{n_e}{\rm cm^{-3}}\right)\left(\dfrac{E}{1\, \rm GeV}\right)^{-1}{\rm Mpc}^{-1}\, ,\\
\Delta_{\rm QED}(E,x_3)&=&\dfrac{\alpha E}{45\pi} \left( \dfrac{B_T(x_3)}{B_{\rm cr}}\right)^2 \simeq4.10\times 10^{-12}  \left(\dfrac{E}{1\, \rm GeV}\right)\left(\dfrac{B_T(x_3)}{1\, \rm n G}\right)^2{\rm Mpc}^{-1}\, ,\\
\Delta_{\rm CMB}(E)&=& \rho_{\rm CMB}E \simeq 0.80 \times 10^{-4}\left(\dfrac{E}{1\, \rm GeV}\right) {\rm Mpc}^{-1}\, ,\\
\Delta_{a\gamma}(x_3)&=&\dfrac{1}{2}g_{a\gamma}B_T(x_3)\simeq 1.52 \times 10^{-2}  \left(\dfrac{g_{a\gamma}}{10^{-11}\, \rm{GeV}^{-1}}\right) \left(\dfrac{B_T(x_3)}{1\, \rm n G}\right){\rm Mpc}^{-1}\, ,~~~~\\
\Delta_{aa}(E)&=&-\dfrac{m_a^2}{2E}\simeq -0.78 \times 10^{2} \left(\dfrac{m_a}{10^{-9}\, \rm GeV}\right)^2\left(\dfrac{E}{1\, \rm GeV}\right)^{-1}{\rm Mpc}^{-1}\, .
\end{eqnarray}
Note that the Faraday rotation terms $\Delta_{12}(E,x_3)$ and $\Delta_{21}(E,x_3)$ can be neglected. 
Here the term $\Delta_{\rm pl}(E,x_3)$ represents the plasma effect when the photon-ALP system propagates in the plasma environment with the plasma frequency 
\begin{eqnarray}
\omega_{\rm pl}= \sqrt{\dfrac{4\pi \alpha n_e}{m_e}}\, ,
\end{eqnarray}
where $\alpha$ is the fine-structure constant, $n_e$ and $m_e$ are the free electron number density and mass, respectively.
The term $\Delta_{\rm QED}(E,x_3)$ represents the QED vacuum polarization effect with the critical magnetic field \cite{Schwinger:1951nm}
\begin{eqnarray}
B_{\rm cr} = \dfrac{m^2_e}{|e|}\simeq4.41\times 10^{13} \rm \, G\, ,
\end{eqnarray}
and the term $\Delta_{\rm CMB}(E,x_3)$ represents the CMB photon dispersion effect with \cite{Dobrynina:2014qba}
\begin{eqnarray}
\rho_{\rm CMB}\simeq0.511\times10^{-42}\, .
\end{eqnarray}
If considering the transversal magnetic field $B_T$ is aligned along the direction $x_2$, the mixing matrix $\mathcal{M}(E,x_3)$ can be rewriten as
\begin{eqnarray}
\mathcal{M}(E,x_3)=
\left(\begin{array}{ccc}
\Delta_{11}(E,x_3)~ &  0&0\\
 0& \Delta_{22}(E,x_3)~ & \Delta_{a \gamma}(x_3)\\
 0&\Delta_{a\gamma}(x_3)  & \Delta_{aa}(E)
\end{array}\right)\, .
\end{eqnarray}
Finally, the photon-ALP conversion probability in the homogeneous magnetic field can be described by
\begin{eqnarray}
\mathcal{P}_{a\gamma}(E,x_3)=\left( \dfrac{g_{a\gamma}B_T L_{\rm osc}(E)}{2\pi}\right)^2 \sin^2\left(\dfrac{\pi x_3}{L_{\rm osc}(E)}\right)\, ,
\end{eqnarray}  
where $L_{\rm osc}(E)$ is the oscillation length
\begin{eqnarray}
\begin{aligned}
L_{\rm osc}(E) &=2\pi\left[\left(\Delta_{22}(E)-\Delta_{aa}(E)\right)^2+4\Delta_{a \gamma}^2\right]^{-1/2}\\
&=2\pi\left[\left[ \dfrac{|m_a^2-\omega_{\rm pl}^2|}{2E}+E\left(\dfrac{7\alpha}{90\pi} \left( \dfrac{B_T}{B_{\rm cr}}\right)^2 +\rho_{\rm CMB}\right)\right]^2+g_{a\gamma}^2B_T^2\right]^{-1/2}  \, .
\end{aligned}
\end{eqnarray}
 
\bibliography{references}

\end{document}